# Si$_3$AlP: A new promising material for solar cell absorber

Ji-Hui Yang[†], Yingteng Zhai[†], Hengrui Liu[†], Hongjun Xiang[†*], Xingao Gong[†*], and Su-Huai Wei[‡]

[†]*Key Laboratory of Computational Physical Sciences (Ministry of Education), State Key Laboratory of Surface Physics, and Department of Physics, Fudan University, Shanghai-200433, P.R. China*

[‡]*National Renewable Energy Laboratory, Golden, Colorado 80401, USA*

Supporting Information Placeholder

**ABSTRACT:** First-principles calculations are performed to study the structural and optoelectronic properties of the newly synthesized nonisovalent and lattice-matched (Si$_2$)$_{0.6}$(AlP)$_{0.4}$ alloy [T. Watkins et al., J. Am. Chem. Soc. **2011**, 133, 16212.] We find that the ordered CC-Si$_3$AlP with a basic unit of one P atom surrounded by three Si atoms and one Al atom is the most stable one within the experimentally observed unit cell.[1] Si$_3$AlP has a larger fundamental band gap and a smaller direct band gap than Si, thus it has much higher absorption in the visible light region. The calculated properties of Si$_3$AlP suggest that it is a promising candidate for improving the performance of the existing Si-based solar cells. The understanding on the stability and band structure engineering obtained in this study is general and can be applied for future study of other nonisovalent and lattice-matched semiconductor alloys.

Recently, significant efforts have been devoted to search new or improve existing photovoltaic materials. Among all the existing solar cell technologies, first-generation crystalline silicon based solar cells are still one of the most important solar cell materials in terms of the energy conversion efficiency and utilization. It has reached an efficiency of more than 23% and has a market share of over 80% in the world photovoltaic industry[2]. However, further development of Si-based solar cells is limited by the intrinsic material properties of Si, i.e., it has an indirect band gap and a relatively low band gap energy (~1.1 eV). For developing new solar cell absorber with optimal performance, it would be desirable to find a new material based on Si, which has a more direct and relatively higher band gap to improve its solar conversion efficiency.

For conventional semiconductors, when group IV elements mutate into its corresponding III-V and II-VI compounds, both the band gap and the directness of the band gap (the energy difference between the fundamental band gap and the lowest direct band gap) increases with the increased ionicity. For example, for Ge, GaAs, and ZnSe, Ge has an indirect band gap of 0.7 eV, whereas GaAs and ZnSe have direct band gaps of 1.5 and 2.8 eV, respectively. Therefore, to modify the band structure of Si for solar absorbers, it will be natural to try alloying Si with its mutated III-V or II-VI semiconductors such as AlP or MgS. These nonisovalent (IV$_2$)$_{1-x}$(III-V)$_x$ alloys are expected to have a wide range of band gaps and more importantly, they are lattice matched, which can provide us great flexibilities to tune the band gap for specific applications such as tandem solar cells. However, only few of these alloys have been studied and applied in practice. One of the examples is BNC$_2$, which is an alloy of diamond and cubic BN. This mixed alloy is chemically more stable than C but also with a super hardness comparable to that of diamond[3-5]. Some studies have also been done on (Ge$_2$)$_x$(GaAs)$_{1-x}$[6-8] alloy for producing an 1 eV direct band gap absorber lattice-matched to GaAs. The main problem for these nonisovalent (IV$_2$)$_{1-x}$(III-V)$_x$ alloys is that under normal growth conditions, these alloys tend to phase separate due to large chemical mismatch, thus, preventing group IV semiconductors from forming homogenous alloys with III-V semiconductors. It is, therefore, very exciting to notice that very recently, Si$_3$AlP, an alloy between Si and AlP, has been successfully synthesized by Watkins et al.[1] using gas source (GS) MBE. This material is lattice matched to Si and could have potential to significantly improve the performance of crystalline Si solar cells. Unfortunately, so far, no theoretical study has been carried out to understand the atomic configurations and optoelectronic properties of this nonisovalent semiconductor alloy.

In this communication, we investigated the stability and optoelectronic properties of Si$_3$AlP and its suitability as a new material for improving the performance of the existing Si-based solar cells by performing first-principles density functional theory (DFT) calculations.[9,10] The local density approximation (LDA) is used to relax the structure parameters as implemented in the VASP code. The electron and core interactions are included using the frozen-core projected augmented wave (PAW) approach[11]. The cutoff kinetic energy for the plane-wave basis wave functions is chosen to be 400 eV for all the calculations. The Brillouin zone (BZ) integration is carried out using 8 x 8 x 6 gamma-centered Monkhorst-Pack k-point meshes[12] for the 10-atoms Si$_3$AlP primitive cell and equivalent k-points for the other cells. All lattice vectors and atomic positions are fully relaxed until the quantum mechanical forces became less than 0.01 eV/Å. For the optical spectrum calculations, we used Bethe-Salpeter equation (BSE) method implemented in the Yambo code, where the LDA band error is corrected using the GW approach.[13] We find that the relative changes of the band structure between Si and Si$_3$AlP is not sensitive to the GW correction.

*Structural analysis of* Si$_3$AlP: Experimental study by Watkins et al.[1] shows that the conventional 20-atom Si$_3$AlP cell consists of four basic units of [AlPSi$_3$], which can be seen as a

$\sqrt{5/2}a \times \sqrt{5/2}a \times a$ Si supercell. In this cell, P atoms are assumed to arrange in a way identical to the C atoms in the Si$_4$C structure[14,15], forming a square lattice with each column separated by a chess Knight move from its neighbors (Fig. 1). We considered the cases where four P atoms in the supercell are fixed, as shown in Figure 1. We found that the formation of Al-Al bonds is not energetic favorable (our test shows that the formation of one Al-Al bond can increase the total energy by as much as 0.59 eV in the 20-atoms Si$_3$AlP cell). By enumerating all the structures without Al-Al bonds, we identified 6 nonequivalent lowest-energy arrangements labeled by their symmetry group in this 20-atom conventional cell as shown in Figure 1, which all keep the basic unit of [AlPSi$_3$].

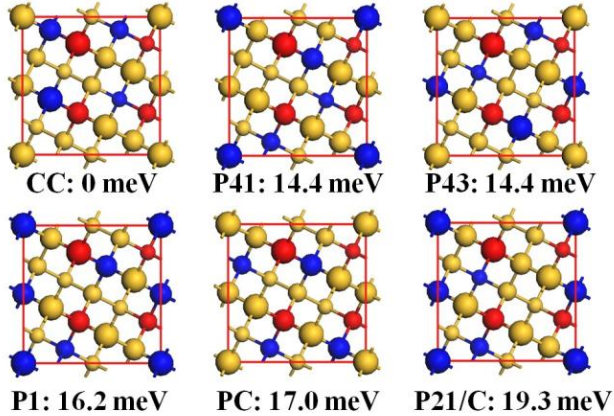

Figure 1. Six different [AlPSi$_3$] atomic arrangements in a 20-atom cell are shown from top of the (001) surface, labeled by their symmetry group and the energy difference per atom relative to CC-Si$_3$AlP. The red balls are P, blue balls are Al and yellow balls are Si. Atoms in the upper layers are drawn with larger sizes and four layers are shown.

After relaxation, the most stable arrangement in this cell adopts a *CC* symmetry structure with 10 atoms in its primitive cell (supporting information (SI)). The relative energies of the other arrangements are also given in Figure 1. One thing to notice is that a more stable structure for Si$_3$AlP is a (110) (Si)$_3$(AlP)$_1$ superlattice, which maximizes the number of the Si-Si and Al-P bonds, and is a precursor of the phase separation[1]. The calculated lattice mismatch between Si$_3$AlP and Si is very small, less than 0.6% (SI), which is consistent with the fact that the experimental lattice constants of Si (5.4306 Å) and AlP (5.4510 Å) are very similar. Another thing to notice is that due to the lower symmetry, the lattice of the ordered *CC*-Si$_3$AlP is a little distorted after relaxation. The angle between a and b is about $89.66°$ instead of the ideal value of $90°$. This distortion could be reduced if the alloys become less ordered. In the following, we will take *CC*-Si$_3$AlP as a representative example to study its electronic and optical properties.

*Electronic band structures of* Si$_3$AlP: We calculated the band structure of *CC*-Si$_3$AlP, which is shown in Figure 2. The valence band maximum (VBM) is at the $\Gamma$ point and the conduction band minimum (CBM) is at a point along the E-K line. Our LDA calculations find that the indirect band gaps of CC-Si$_3$AlP is 0.16 eV larger than that of Si, so adding AlP to Si increases the fundamental band gap. For CC-Si$_3$AlP the minimum direct band gap is found at a point along the $\Gamma$-A line (Figure 2). The calculated direct band gap at the $\Gamma$ point for CC-Si$_3$AlP is 0.33 eV smaller than that of Si at the $\Gamma$ point, thus adding AlP to Si reduces the optical band gap. The increase of the indirect fundamental band gap of Si$_3$AlP than Si is beneficial in increasing the open-circuit voltage and the decrease of the direct optical band gap is beneficial in increasing the absorption, thus, the photocurrent of the solar cell. The calculated band structures for other configurations shown in Figure 1 are quite similar to the CC structure, with variation of the fundamental band gap less than 0.12 eV, indicating that as long as the basic motif is the same, the indirect band gap is not sensitive to the atomic arrangements.

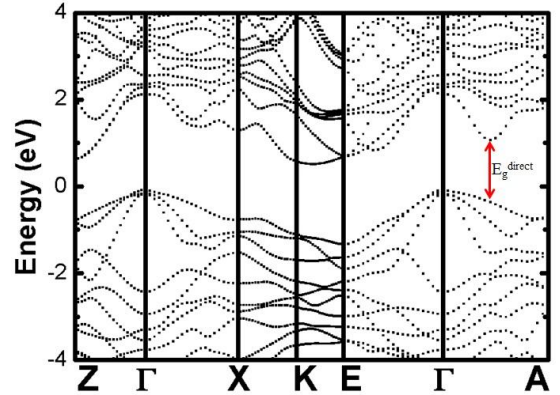

Figure 2. Calculated LDA band structures of CC-Si$_3$AlP.

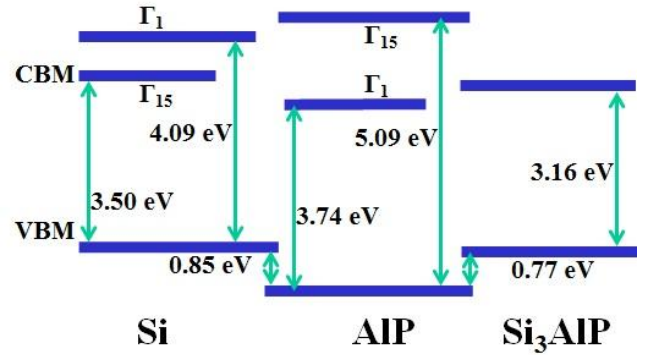

Figure 3. Band alignments at $\Gamma$ point between Si, AlP, and Si$_3$AlP. The VBM states are aligned through our band offsets calculations while the CBM states are aligned using the experimental band gaps of Si and AlP and GW band gap corrections.

The change of the optical band gap can be understood as follows. For the minimum direct band gap along the $\Gamma$-A line, the VBM at this k point originates from the Si 3p state, whereas the CBM is derived mostly from Si 3s state. The transition between the two states is not allowed in pure Si because they are folded in from two different k-points in the 2-atom Si BZ. But after AlP is mixed into Si, due to the reduced symmetry, non-diamond potential is introduced to couple different folded states and this transition becomes allowed for Si$_3$AlP. For the direct band gap at $\Gamma$ point, before AlP is introduced, the CBM originates from the Si 3p state. After AlP is mixed, the CBM energy at $\Gamma$ is expected to initially increase and switch to a more s-like state because AlP p-like $\Gamma_{15}$ conduction band state is much higher in energy than the s-like $\Gamma_1$ state (Figure 3). This, however, contradicts to the calculated results which show the CBM at $\Gamma$ decreases when AlP is added. This is because in the supercell of Si$_3$AlP, there are states folded to the $\Gamma$ point and these states will couple with the CBM state at $\Gamma$ and push the CBM down. These two effects lead to the lowering of the CBM of *CC*-Si$_3$AlP than that of Si. Similar case also happens to the VBM state. On one hand, the introduction of AlP will lower the VBM state; on the other hand, coupling to the folded states at $\Gamma$ point pushes the VBM upward. So the VBM of *CC*-Si$_3$AlP is only a little lower than that of Si. Taking all the above into account, the optical band gap at $\Gamma$ point of *CC*-Si$_3$AlP is smaller than that of Si or AlP, which is beneficial to generate more photocurrent.

*Optical spectrum of* Si$_3$AlP: The calculated absorption coefficients as a function of energy are shown in Figure 4. The GW plus BSE result of Si agrees very well with the experimental result.[16] From the calculated results, we can see that, relative to the absorption spectrum of Si, Si$_3$AlP has a much higher absorption in the low energy region, which is caused by its smaller optically active direct band gap than that of Si. The GW+BSE results show that the sharp jump of the optical spectrum is at about 3.0 eV in Si$_3$AlP, compared to about 3.2 eV in Si, which are close to their respective direct band gaps at the $\Gamma$ point. However, for Si$_3$AlP**,** the absorption below this sharp increase is also significantly increased due to the alloying induced/enhanced optical absorption. We also calculated the optical spectrums of the other structures shown in Figure 1 (see SI). We find that they have almost the same optical spectrums as the CC structure. This indicates that the optical properties of these materials are not sensitive to the atomic arrangements in the unit cell, which provides convenience for the synthesis.

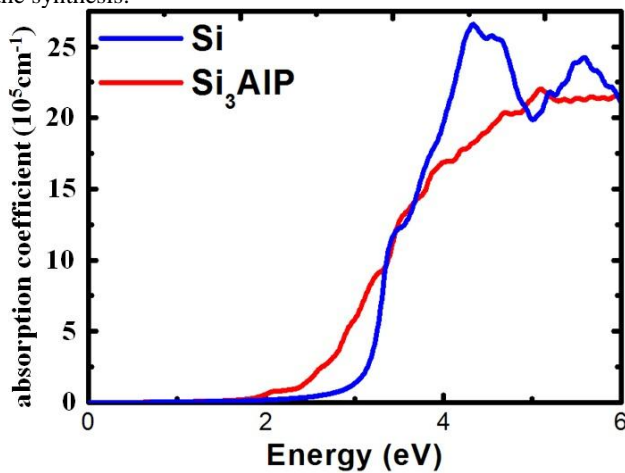

Figure 4. Calculated diagrams of absorption coefficient as a function of energy using the GW+BSE method for Si and CC-Si$_3$AlP.

In conclusion, we have systematically studied the structural, electronic and optical properties of the newly synthesized nonisovalent and lattice matched (IV$_2$)$_{1-x}$(III-V)$_x$ alloy Si$_3$AlP. The CC-Si$_3$AlP is found to be the most stable structure within the experimentally observed unit cell, but the electronic and optical properties of the alloy is not sensitive to the atomic arrangement in the unit cell as long as the local [Si$_3$AlP] clusters are maintained. We find that Si$_3$AlP has a larger fundamental band gap and a smaller direct band gap at $\Gamma$ than Si, thus is more suitable for solar cell absorbers than Si. Therefore, we propose that Si$_3$AlP could be a strong candidate for photovoltaic applications. Experimental efforts for studying this material for photovoltaic absorber are called for.

## ASSOCIATED CONTENT

Supporting Information. **Primitive cell of CC-Si$_3$AlP, lattice parameters of relaxed structures of Si, AlP and Si$_3$AlP using LDA functional, band structures of Si in its 2-atoms cell and in 10-atoms CC- Si$_3$AlP cell, and absorption coefficients versus energy for all the 6 different structures. This material is available free of charge via the Internet at http://pubs.acs.org.**

## AUTHOR INFORMATION

### Corresponding Author

hxiang@fudan.edu.cn; xggong@fudan.edu.cn

## ACKNOWLEDGMENT

The work at Fudan University is partially supported by International collaboration project, NSFC, Special funds for major state basic research, Pujiang plan, and Program for Professor of Special Appointment (Eastern Scholar). The computation is performed in the Supercomputer Center of Fudan University. The work at NREL is funded by the U.S Department of Energy (DOE), under Contract No. DE-AC36-08GO28308.

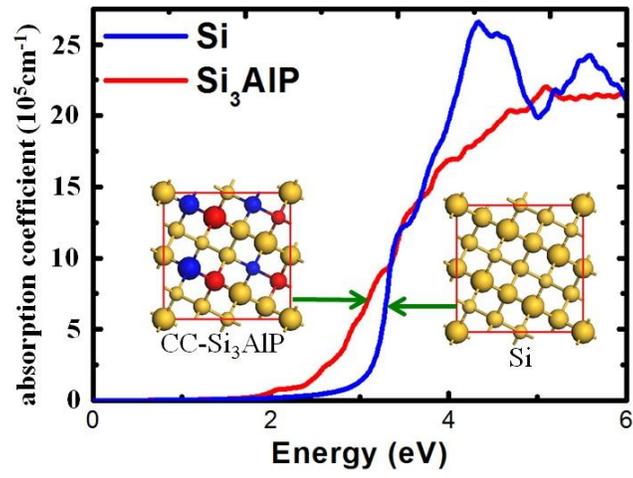

Toc Graphic